\documentclass[12pt,aps,amsmath,latexsym,amsfonts,letterpaper]{JHEP3}
\usepackage[all]{xy}

\usepackage{epsfig}
\usepackage{subfigure}

\def\real{\mathbb R}

\def\half{\textstyle{1\over2}}

\def\ie{{\it i.e.,}\ }

\newcommand{\be}{\begin{equation}}
\newcommand{\ee}{\end{equation}}
\newcommand{\bea}{\begin{eqnarray}}
\newcommand{\eea}{\end{eqnarray}}
\newcommand{\bml}{\begin{mathletters}}
\newcommand{\eml}{\end{mathletters}}

\newcommand{\ga}{\ensuremath{\gamma}}

\newcommand{\mn}{\ensuremath{{\mu\nu}}}

\newcommand{\eps}{\ensuremath{\epsilon}}
\newcommand{\s}{\ensuremath{\sigma}}

\renewcommand{\tt}{\ensuremath{\textrm}}
\renewcommand{\t}{\ensuremath{\tau}}

\newcommand{\del}{\ensuremath{\partial}}

\newcommand{\m}{\mathcal}
\newcommand{\ba}{\begin{eqnarray}}
\newcommand{\ea}{\end{eqnarray}}

\newcommand{\lab}{\label}
\newcommand{\beq}{\begin{equation}}
\newcommand{\eeq}{\end{equation}}
\newcommand{\beqa}{\begin{eqnarray}}
\newcommand{\eeqa}{\end{eqnarray}}
\newcommand{\beqar}{\begin{eqnarray*}}
\newcommand{\eeqar}{\end{eqnarray*}}
\newcommand{\reef}{\ref}

\newcommand{\eg}{{\it e.g.,}\ }

\renewcommand{\a}{{a}}
\newcommand{\B}{{\Delta S}}

\title{A New Perspective on DGP Gravity}
\author{Ruth Gregory,$^a$ Nemanja Kaloper,$^b$ Robert C. Myers$^c$
and Antonio Padilla$^d$\\
$^a$ Centre for Particle Theory, Durham University, South Road,
Durham,\\ \ \ DH1 3LE, UK\\
$^b$ Department of Physics, University of California, Davis, CA
95616, USA\\
$^c$ Perimeter Institute for Theoretical Physics,
Waterloo, ON, N2L 2Y5, Canada \\
$^c$ Department of Physics and Astronomy, University of Waterloo,
Waterloo,\\ \ \  ON,  N2L 3G1, Canada \\
$^c$ Kavli Institute for Theoretical Physics, University of
California,\\ \ \ Santa Barbara, CA 93106-4030, USA\\
$^d$ School of Physics and Astronomy, University Park, University of
Nottingham,\\ \ \ Nottingham NG7 2RD, UK}

\date{\today}

\abstract{We examine brane induced gravity on codimension-1 branes,
{\it a.k.a} DGP gravity, as a theory of five-dimensional gravity containing
a certain class of four-dimensional branes. From this perspective,
the model suffers from a number of pathologies which went unnoticed
before. By generalizing the $5D$ geometry from Minkowski to Schwarzschild, 
we find that when the bulk mass is large enough, 
the brane hits a pressure singularity at finite radius. Further, 
on the self-accelerating branch, the five-dimensional energy is 
unbounded from below, implying that the self-accelerating backgrounds are 
unstable. Even in an empty Minkowski bulk, standard Euclidean
techniques suggest that the spontaneous nucleation of
self-accelerating branes is unsuppressed. If so, 
quantum effects will strongly modify any classical
intuition about the theory. We also note that unless considered as
$\mathbb{Z}_2$-orbifold boundaries, self-accelerating branes
correspond to `wormhole' configurations, which introduces the usual
problematic issues associated with wormholes. Altogether these pathologies
present a serious challenge that any proposed UV completion of the
DGP model must overcome.}

\keywords{\it brane worlds, modified gravity, cosmic acceleration, ghosts}
\preprint{arXiv:0707.2666 [hep-th]\\ DCPT-07/33}

\begin{document}

\section{Introduction}

The observed late-time acceleration of our universe \cite{snova}
presents an enormous puzzle for theoretical physicists. There has
been a great deal of effort invested in the search for a
self-consistent modification of gravity as a way to address this
problem, instead of introducing new matter contributions to Einstein
gravity such as the cosmological constant. Perhaps the most notable
examples of this approach are braneworld modifications
\cite{GRS,DGP,ASY, CGP} and theories with new scalars disguised as
$f(R)$ terms \cite{cap}. The brane induced gravity model in $5D$
(henceforth refered to as DGP gravity, for short) \cite{DGP} in
particular has received special attention, mainly because it gave
rise to the {\it self}-accelerating (SA) backgrounds. These
solutions describe de Sitter cosmology without a nominal
cosmological constant, or tension, on the brane, as have been
written down in \cite{Cedef,Cedef2}. They are similar in spirit to
the original inflationary models of Starobinsky, found in higher
derivative gravity without cosmological constant before the full
power of inflationary dynamics was realized \cite{staro}. At first,
the SA solutions appeared to evade any inconsistencies and in
particular, the ghost problem of the GRS model
\cite{GRSghost,piraza}. However, subsequent investigations
demonstrated that a ghost appears on the SA brane
\cite{lupora,bend}. The appearance of both ghosts and tachyons was
confirmed by the careful analysis of fluctuations around the SA
solutions in \cite{CGKP,KKK}, which also showed that the normal
branch of solutions remain free of ghosts and tachyons. While the
appearance of ghosts and tachyons would seem to present a major
problem for the DGP model, differing views have emerged as to the
severity of this pathology \cite{non,dispute} -- in particular, it
is has been argued that strong coupling phenomena may somehow alter the
naive picture of instabilities. On the other hand, the exact shock
wave analysis revealed some pathological singularities on the SA
branch of solutions even beyond perturbation theory \cite{shocks}.
It is therefore clear that the issues concerning ghosts and tachyons
remain a major obstacle to working with the DGP model as a reliable
and practical description of our universe.

Most discussions of the DGP gravity to date focussed on the
four-dimensional physics that brane observers might experience. Here
we take a different point of view and examine the DGP model as a
theory of five-dimensional Einstein gravity coupled to an unusual
source: the four-dimensional DGP branes. In developing this
perspective we uncover a number of new pathologies, which went
unnoticed before. 

The first of these are seen by generalizing the $5D$ geometry 
from Minkowski to Schwarzschild. If the bulk mass exceeds 
a certain critical value, the brane will run into a pressure singularity 
at finite radius. Further, it is straightforward to see that on the SA branch the
five-dimensional energy is unbounded from below. Because an SA brane excises the
space where the naked singularity would have been, the remaining spacetime 
of negative $5D$ mass is nonsingular, and we can keep all these solutions 
in the spectrum of the theory. This reservoir of
negative energies can be accessed by smooth deformations of the SA
vacua, by turning on the radion field which encodes the bulk ADM
mass. While such deformations are not immediately accessible to the
gravitational modes on the normal branch backgrounds alone, since
there is no normalizable radion in that case, backgrounds which
contain any number of SA branes will reintroduce this instability.

Going back to a Minkowski bulk, we can use the 
standard semiclassical techniques in Euclidean quantum
gravity to demonstrate that the mixing between empty $5D$ 
space and SA branes appears to be completely unsuppressed. This
result would lead us to conclude that as long as the theory contains
SA brane solutions, $5D$ Minkowski space cannot be a good
approximation of the (quantum) vacuum of the theory. Of course, 
these calculations in the Euclidean path integral
are subject to certain ambiguities -- as we will discuss. Finally,
we note that unless the DGP branes are considered to be
$\mathbb{Z}_2$-orbifold boundaries, the self-accelerating branes are
actually `wormhole' configurations. This will automatically
introduce the usual set of problematic issues associated with such
geometries \cite{worm}.

Our results complement the perturbative exploration of pathologies
in the DGP model from the viewpoint of brane-localized observers,
which revealed ghosts and tachyons in the spectrum of small
perturbations about the SA solutions \cite{CGKP,KKK,joel,morepert}.
We uncover the present pathologies using the nonperturbative
classical theory, which probes the full nonlinearities of the
gravitational theory, in contrast to the perturbative analysis of
\cite{CGKP,KKK}. This shows that the nonlinearities, on their own,
do not remedy the perturbative maladies of the model. The present
calculations are extremely simple, utilizing the high degree of
symmetry expected to characterize physically relevant backgrounds
for cosmology at the Hubble scales. This simplifies the technical
aspects of the work in contrast to the general perturbative
considerations of \cite{CGKP,KKK}. The results which we obtain
provide clear and simple
examples of how strong coupling effects {\it do not} cure
sub-crossover dynamics on their own. 
Indeed, irrespective of any strong coupling phenomena,
the solutions with arbitrarily negative energies remain accessible
to classical dynamics, at least in some particular dynamical channels. 
On the other hand, this also
affirms that in its present form the low energy description better
be very sensitive to higher-derivative, strong coupling corrections
if one wants to maintain the belief that such phenomena may alter our
conclusions. We stress that while the pathologies found here paint a disturbing
picture of the DGP model, we cannot rule out the possibility that
this model may have a sensible UV completion. It appears clear,
however, that these pathologies do present a severe challenge for
any such proposed UV completion. While the model at its current
stage may motivate the search for a complete theory, given the
severity of the pathologies revealed here as well as the presence of
tachyonic and ghost-like perturbations, it seems premature to use
its background solutions, and in particular the self-accelerating
cosmologies, as a basis for constructing any detailed observational
tests, such as those pursued in \cite{zalda,lueglen} and many
subsequent works. Moreover, our results seem to identify the
self-accelerating branes as the source of instability. Thus this
suggests that if stable UV completions of the DGP model should
exist, they may not allow for self-accelerating branes, altogether
invalidating the utility of such solutions. We note that the
perspective and calculations developed here are applicable more
generally in a variety of other settings and may be used as a
diagnostic in other higher dimensional models or braneworld
scenarios, of interest as a modified theory of gravity.

We should mention here a recent interesting work \cite{kyony},
where the authors sought configurations describing transitions from SA branes
to N branch solutions, that look like normal branch bubbles
on the SA brane. Such solutions require domain walls separating different
de Sitter phases, and the authors of \cite{kyony} argued that smooth walls supported by
positive definite field theories may not exist. On this ground, these authors suggested
that perhaps the perturbative ghost of DGP model might be less ominous than one
may think. Now, while these arguments may be questioned, we should stress that our arguments
are completely orthogonal to the investigation of \cite{kyony}, and show that nonperturbative
instabilities {\it do exist}. Hence, at the level of the semiclassical theory,
the ghost of DGP is not tamed, but comes out with a vengeance.

The paper is organized as follows: section \ref{model} provides
a brief overview of the DGP model. In section \ref{negative}, we
examine the various cosmological solutions of this model where the
bulk spacetime corresponds to the five-dimensional Schwarzschild
solution. For a large enough (positive) mass in the bulk, 
we show that the brane runs into a pressure singularity at finite radius. 
We further point out that on the SA branch, 
the negative-mass Schwarzschild subfamily still represent a class of smooth solutions 
of the DGP theory, yielding states which have a negative energy from the
five-dimensional perspective. In section \ref{mix}, we turn to
semiclassical calculations describing the nucleation of a
self-accelerating brane in empty $5D$ Minkowski space.
We close with a discussion of our results in section \ref{discuss}.
Appendix \ref{radion} examines the solutions of section
\ref{negative} in the limit where the Schwarzschild mass parameter
is small and connects this limit with the perturbative analysis of
\cite{CGKP}.

\section{The DGP Model}\lab{model}

The $5D$ DGP model, as we study it here, is described by the following
action:
\be
%\fl
S=2M_5^3\int_\textrm{bulk} \sqrt{-g} R+4M_5^3\int_\textrm{brane}
\sqrt{-\gamma} K+\int_\textrm{brane} \sqrt{-\gamma}\left(M_4^2
\mathcal{R}-\sigma+\mathcal{L}_{matter}\right)
\lab{action}\ee
where $g_{ab}$ is the bulk metric with corresponding Ricci tensor
$R_{ab}$. The brane has induced metric $\gamma_\mn$ with
corresponding Ricci tensor $\mathcal{R}_\mn$. Its  extrinsic
curvature is given by $K_\mn=-\frac{1}{2}{\cal L}_n \ga_\mn$, the Lie
derivative of the induced metric, with respect to the unit normal
$n^a$, pointing {\it into} the bulk. For the most part, we will work with
the solutions which are $\mathbb{Z}_2$-symmetric about the
brane. This may be thought as a simplifying constraint which focuses
our attention on a special class of solutions. Alternatively, it may
be that the brane is actually a $\mathbb{Z}_2$-orbifold in which
case we would think of this as a boundary of the five-dimensional
spacetime. On a pragmatic level, the $\mathbb{Z}_2$-symmetry means
we only ever work with one side of the bulk in the following.

The key feature of the DGP model is the intrinsic curvature term
appearing on the brane. In general, one might think that such a term
should be induced on the brane by matter loop
corrections~\cite{loops,colho,lowe} or finite width
effects~\cite{width}. However, as we comment below, the
phenomenologically interesting case requires a hierarchically much
larger brane Planck scale $M_4$ than the bulk Planck scale
$M_5$~\cite{DGP}, which has not been easy to realize naturally, and
has been questioned on theoretical grounds \cite{noUV}. In the brane
action, we have explicitly extracted the brane tension $\sigma$ out
of the matter Lagrangian $\m{L}_{matter}$. The tension term can be
viewed as encoding the vacuum energy of the brane-localized fields.
Then, the governing equations of motion in the bulk are the vacuum
Einstein equations
\be G_{ab}=R_{ab}-\half R g_{ab}=0\,. \lab{bulk} \ee
The boundary conditions at the brane are simply the Israel junction
conditions extended to the present case:
\be
4M_5^3K_\mn+2M_4^2 \left(\m{R}_\mn-\frac{1}{6} \m{R}\ga_\mn
\right)-\frac{\sigma}{3}\ga_\mn=
T_\mn-\frac{1}{3}T\,\ga_\mn\lab{israel} \ee
where $T_\mn\equiv-\frac{2}{\sqrt{-\ga}}\frac{\del\left(\sqrt{-\ga}
\m{L}_{matter}\right)}{\del \ga^\mn}$. We include it here for
generality, although it will vanish for the solutions which we will consider below.

There are a variety of different approaches to solving the coupled
equations (\ref{bulk},\ref{israel}). Here we will first solve the
bulk equations and then use the Israel junction conditions to
determine the trajectory of the brane in this bulk. For our
purposes, the first step amounts to choosing a known solution of the
five-dimensional vacuum Einstein equations. Of course, the simplest
of these is just five-dimensional Minkowski space:
\begin{equation}
ds^2=g_{ab}dX^a dX^b=-dt^2+dr^2+r^2d\Omega^2_3\,.\lab{minkowski}
\end{equation}
The brane sweeps out a surface in this background at which two
copies of the bulk geometry are stitched together (or an orbifold
boundary condition is imposed). A simple but interesting brane
geometry to consider is a homogeneous cosmology
\begin{equation}
ds^2=\gamma_\mn dx^\mu dx^\nu=-d\tau^2 +\a(\tau)^2 \, d
\Omega_3^2\,,\lab{cosmogeo}
\end{equation}
where $\a(\tau)$ specifies the proper size of the brane as a
function of its proper time $\tau$. To specify the brane's
trajectory or embedding in the background (\reef{minkowski}), we set
$r=\a(\t)$. Omitting the details of analysing the Israel conditions
(\reef{israel}) for this case here\footnote{It corresponds to the
special case $\mu=0$ of the calculation in the next section.} we
just note that they yield the following interesting vacuum solutions
on the brane~\cite{Cedef,Cedef2,kalin}
\begin{equation}
\a(\tau)=\frac{1}{H_\pm}
  \cosh(H_\pm \tau) \lab{cosmosol}
\end{equation}
where
\be H_\pm\equiv\half H_0 \left(\sqrt{1+\frac{4\s^*}{H_0}}\pm
1\right)\,. \lab{H} \ee
Here we have introduced $\sigma^*=\sigma/12M_5^3$ and
$H_0=2M_5^3/M_4^2$. These solutions describe de Sitter geometry on
the brane with radius of curvature $H^{-1}$. As our construction
elucidates, the brane can be viewed as a $4D$ hyperboloid of the
same radius embedded in the $5D$ Minkowski bulk (see figure
\ref{fig:one}), generalizing the inflating domain walls of
\cite{vis}. The choice of sign appearing in (\reef{H}) arises
because the construction left ambiguous which part of the bulk
spactime was included. Indeed, in the minimal approach, we can treat
the brane as a $\mathbb{Z}_2$ orbifold \cite{DGP}, in which case we
identify the different sides of the bulk. Then the `minus' sign
denotes pasting together two copies of the region interior to
the hyperboloid, and the `plus' sign refers to pasting two copies of the
exterior.
\FIGURE{\includegraphics[width=10cm, height=9cm]{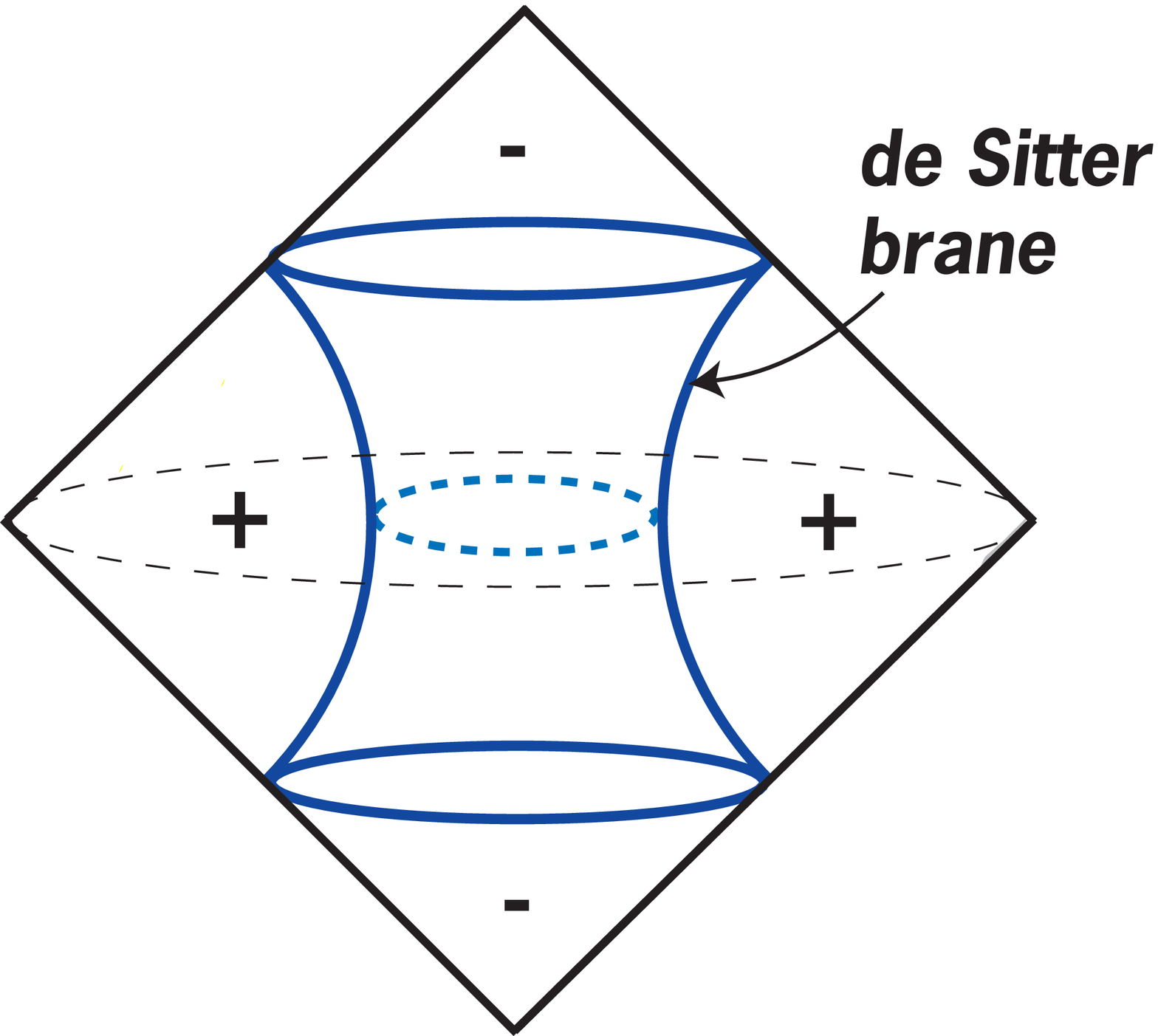}
\caption{Embedding of a de Sitter brane in a flat 5D bulk.
The brane world volume is the hyperboloid in the Minkowski bulk. The
normal branch corresponds to keeping the interior of the
hyperboloid, the self-accelerating branch the exterior.} \lab{fig:one}}

The solution with $H_-$ is commonly referred to as the {\it normal}
branch whereas the solution with $H_+$ is referred to as the {\it
self-accelerating} branch, a terminology which will become
transparent shortly. The value of $H_0=2M_5^3/M_4^2$ in (\reef{H})
is typically taken to be the current Hubble scale. This illustrates
vividly the need for an exponential hierarchy between the Planck
scales: given that $M_4$ is set to the observed four-dimensional
Planck scale, $M_4 \sim 10^{19} \, {\rm GeV}$, to get $H_0 \sim
10^{-33} \, {\rm eV}$, one must take $M_5 \sim 10^{-3} \, {\rm eV}$,
some thirty orders of magnitude smaller than $M_4$. One must explain
how to generate, and stabilize against radiative corrections, such a
disparate ratio between the two Planck scales.

On the other hand, a new and very interesting feature of the theory is
that even for vanishing tension, the self-accelerating solution gives
rise to a de Sitter
brane universe with $H=H_0$. The modification of gravity at large
distances enables us to describe an accelerating universe in the
absence of any vacuum energy whatsoever! In contrast, the normal
branch  gives rise to a flat Minkowski brane as $\s \to 0$,
which may be less interesting for the phenomenology
of an accelerating universe,
but may still be a useful testing ground for other effects of gravity
modified at large scales.

Returning to our geometrical picture of the branes as hyperboloids
in Minkowski space, we note that the self-accelerating solution
leads to a rather counter-intuitive result. Recall that in that case
we keep the {\it exterior} of the hyperboloid. In the absence of the
induced curvature of the brane, this would be consistent with a
brane of {\it negative} tension. But this is not the case here. What
has happened is that the induced curvature term enables us to mimic
negative tension even when $\sigma>0$. If we cast the brane
equations (\reef{israel}) in the conventional form of the Israel
junction conditions, we would have
\beq 4M_5^3\left(K_\mn-K\,\ga_\mn\right)=T^{eff}_\mn\equiv
-2M_4^2\left(\m{R}_\mn-\frac{1}{2} \m{R}\ga_\mn
\right)-\sigma\,\ga_\mn+T_\mn\lab{stress}\eeq
where the effective stress tensor for the brane comes from the
variation of the total brane action, \ie the third integral in
eq.~(\reef{action}). Hence the right-hand side of the Israel
conditions include the standard contributions coming from the brane
tension and the brane matter, both of which will satisfy the usual
positive energy conditions \cite{usual} (with positive $\sigma$).
However, there is also a geometric contribution, \ie the Einstein
tensor of the intrinsic metric, which in general will not satisfy
any positivity conditions. Note the overall minus sign in front of
this term which arises here because we have put this geometric term
on the matter side ({\it the wrong side}) of the equations in
eq.~(\reef{stress}). In particular, for a brane with a de Sitter
geometry, as considered above, this term contributes
$6\,M_4^2H^2\,\ga_\mn$ to the effective stress tensor, \ie this
geometric contribution is equivalent to a negative brane tension
$\sigma_\textrm{geometry}\equiv-6\,M_4^2H^2$. This term comes to
dominate on the self-accelerating branch, for which we may think in
terms of the total effective tension being negative,
$\sigma_\textrm{eff}=\sigma-6\,M_4^2H_+^2=-12M_5^3 H_+$.

As commented above, the effective stress tensor will not in general
satisfy any of the energy conditions typically considered in
Einstein gravity \cite{usual} because of the geometric term
appearing on the right-hand side of eq.~(\reef{stress}). Hence one
must worry that unusual or problematic results may arise from the
brane and/or excitations of the brane. In particular, excitations of
the brane geometry, \eg four-dimensional gravitons, may contribute
negative energies from a five-dimensional perspective. To leading
order, this contribution might be expected to vanish as the
gravitons would satisfy (something-like) the four-dimensional
Einstein equations and so their leading contribution to
$T^{eff}_\mn$ might then vanish. Of course, this question requires a
detailed analysis to take into full account the couplings between
fields on the brane and in the bulk. However, this is precisely the
analysis provided in \cite{CGKP, review} where it was found that
certain perturbations about the self-accelerating solution are
indeed ghost-like, as well as finding other tachyonic modes.

We close here with one final observation. If we relax the
$\mathbb{Z}_2$ boundary conditions on the brane, then the
conventional interpretation of the SA branch has the brane
connecting two asymptotically flat regions of space. Hence this
configuration provides what is widely known in the relativity
literature as a `wormhole'. In particular, while implicitly the two
sides of the SA brane are taken to be disjoint Minkowski spaces,
there is no reason a priori why these branes should not be
connecting distant locations within the same five-dimensional space,
which makes the brane's role as a `wormhole' more evident. Hence we
recall that, as has been extensively considered, there is a set of
problematic issues inherent to such wormhole configurations, such as
the appearance of closed time-like curves -- see, for example,
\cite{worm}. We do not consider these issues further here but the
wormhole geometry will have interesting implications in the context
of the semiclassical calculations in section \ref{mix}. Of course,
these problems are eliminated if the DGP branes are actually defined
to be $\mathbb{Z}_2$-orbifolds. In such a case, we would think of
the brane as a boundary of the five-dimensional spacetime and there
would be no (independent) geometry on the other side.

\section{Branes in a Schwarzschild bulk}\lab{negative}

\subsection{Finding solutions}
 
The cosmological solutions above were based
on the simplest possible bulk, namely,
$5D$ Minkowski space (\reef{minkowski}). It is
straightforward to extend this procedure to other
bulk solutions with spherical symmetry.
In what follows, we apply this procedure to a more interesting bulk
solution, a five-dimensional Schwarzschild black hole:
\begin{equation}
ds^2=g_{ab}dX^a
dX^b=-f(r)dt^2+\frac{dr^2}{f(r)}+r^2d\Omega^2_3\,,\lab{schwarz}
\end{equation}
where
\begin{equation}
f(r)=1-\frac{\mu}{r^2}\,.\lab{schwarz2}
\end{equation}
Given the $5D$ action (\reef{action}), the mass is given by
\cite{mp}
\be m= 12\pi^2\,M_5^3\,\mu \,.\lab{mass}\ee
In the context of DGP gravity, the possibility of using this bulk
geometry was considered in passing in \cite{Cedef}.
%nk proposed change
We will examine the corresponding solutions in full detail. A key
observation will be that there is no obstacle to
constructing smooth solutions on the SA branch when the mass (\reef{mass}) is arbitrarily 
negative, because the SA brane excises the naked singularity. 
We note that the authors of \cite{lupora} considered in more detail a special subfamily                                                 
of SA branes in negative mass bulks. By demanding that the brane                                                           
remains static they inferred a lower bound on the mass. However such a                                                     
bound is artificial as it only implies that for a more negative bulk mass                                                  
the brane with a fixed tension can't be static, but it crunches into a singularity, as we                                                       
will discuss below. 
%nk end change

Let us adopt the same ansatz for the brane geometry and trajectory as above
\begin{equation}
ds^2=\gamma_\mn dx^\mu dx^\nu=-d\tau^2 +\a(\tau)^2 \, d
\Omega_3^2\,.\lab{cosmogeo2}
\end{equation}
with $r=\a(\t)$. To evaluate the Israel junction conditions
(\reef{israel}), we must first examine the embedding geometry more
carefully. As well as the tangent vectors along the three-sphere,
there is a (future-pointing) time-like tangent vector on the brane
which may be expressed as\footnote{Implicitly here and in the
following, we assume that $f(\a)>0$, \ie the brane is outside of the
black hole horizon.}
\be u^a=\left(\frac{dt}{d\t},\frac{dr}{d\t},0,0,0\right)
=\left(\frac{1}{f(\a)}\left(f(\a)+\dot\a^2\right)^{1/2},\dot\a
,0,0,0\right)\,, \lab{tangent}\ee
normalized such that $u\cdot u=-1$. The normal to the brane is
given by
\be n^a=\epsilon\,
\left(\frac{\dot\a}{f(\a)},\left(f(\a)+\dot\a^2\right)^{1/2},0,0,0\right)
\,, \lab{normal}\ee
with $n\cdot n=+1$ and $u\cdot n=0$. Here $\epsilon=\pm1$, which
corresponds to the ambiguity of which part of the bulk is included
in the construction, noted above. With $\epsilon=-1$ ($+1$), $n^a$
points towards the black hole (asymptotic infinity) and we keep the
interior (exterior) region.

Now we can evaluate each of the contributions in
(\reef{israel}). One finds that the extrinsic curvature is
\be K_{ij}=-\half {\cal L}_n
\ga_{ij}=-\frac{\epsilon}{\a}\left(f(a)+\dot\a^2\right)^{1/2}\,\ga_{ij}\lab{extr}
\ee
where $(i,j)$ indicate directions on the $S^3$. With the brane geometry
(\reef{cosmogeo2}), one easily evaluates the intrinsic curvatures as
\be \m{R}_{ij}-\frac{1}{6} \m{R}\ga_{ij}=
\left(\frac{\dot\a^2}{\a^2}+\frac{1}{\a^2}\right) \,\ga_{ij}\,.
\lab{intr}\ee
Evaluating (\reef{israel}) for the three-sphere directions yields
an expression which we write as
\be
-\epsilon\,H_0\,\left(\frac{\dot\a^2}{\a^2}+\frac{f(\a)}{\a^2}\right)^{1/2}
+ \frac{\dot\a^2}{\a^2}+\frac{1}{\a^2}-H_0\s^*=0\,. \lab{israel2}\ee
To understand this result better, we first re-express it in a form
reminiscent of the Friedmann equation
\beqa \frac{\dot\a^2}{\a^2}+\frac{1}{\a^2}&=&\frac{H_0^2}{2}+H_0\s^*
+\epsilon \frac{H_0^2}{2}
\left(1+\frac{4\s^*}{H_0}-\frac{4\mu}{H_0^2}\frac{1}{\a^4}\right)^{1/2}
\lab{expand}\\
{}_{\a\rightarrow\infty}&\simeq&\frac{H_0^2}{2}+H_0\s^* +\epsilon
\frac{H_0^2}{2} \left(1+\frac{4\s^*}{H_0}\right)^{1/2}
-\frac{\epsilon\,\mu}{\a^4}\left(1+\frac{4\s^*}{H_0}\right)^{-1/2}+\ldots
\nonumber\eeqa
where we have used (\reef{schwarz2}). In the second line, we have
expanded the right-hand side for large $\a$ and we can interpret the
result in terms of the standard four-dimensional Friedmann equation.
The first contribution corresponds to that of a cosmological
constant. We recognize this expression as $H_+^2$ for $\epsilon=+1$
and $H_-^2$ for $\epsilon=-1$. The next contribution with a $1/\a^4$
dependence matches that of a radiation gas with a density
proportional to $\epsilon\,\mu$ -- an effective `dark radiation'
that is commonplace in braneworld cosmology \cite{holo}. Note that
this `holographic' effect has an unusual character for the SA brane.
While a positive $\mu$ induces a positive energy density on the N
brane, the effective energy density is negative on the SA brane~\cite{tetra}.
That is, the holographic or four-dimensional interpretation might be
that the SA brane supports some additional ghost-like matter in the
$\mu>0$ background. Of course, perhaps this is simply another
example of the often noted result that the bulk contributions to the
effective stress-energy on the brane \cite{bstress,csaki,marcopel}
may be negative. A nonvanishing $\mu$ also induces an infinite
series of higher order terms proportional to $1/\a^{4n}$ -- many of
which would be expected to be negative.

Setting $\mu=0$, we recover the simple case examined in the previous
section and one can easily verify that eqs.~(\reef{cosmosol}) and
(\reef{H}) give a solution of (\reef{expand}). In general, we have no
analytic solutions for (\reef{expand}). However to gain intuition for
the solutions we can rewrite the equation as the Hamiltonian
constraint of a classical point particle,
\be \dot\a^2+U(\a,\eps,\mu)=-1\,,\lab{classic}\ee
moving in an unusual potential:
\beq U(a,\eps,\mu)=-\frac{H_0^2}{2}
\a^2\left(1+\frac{2\s^*}{H_0}+\epsilon\left(1+\frac{4\s^*}{H_0}-\frac{4\mu}{H_0^2}\frac{1}{\a^4}
\right)^{1/2}\right)\,. \lab{pot}\eeq
The effective potential $U(\a,\eps,\mu)$ is plotted in figure
\ref{fig:pot} for a variety of parameter choices.
\FIGURE{\includegraphics[width=11cm, height=9cm]{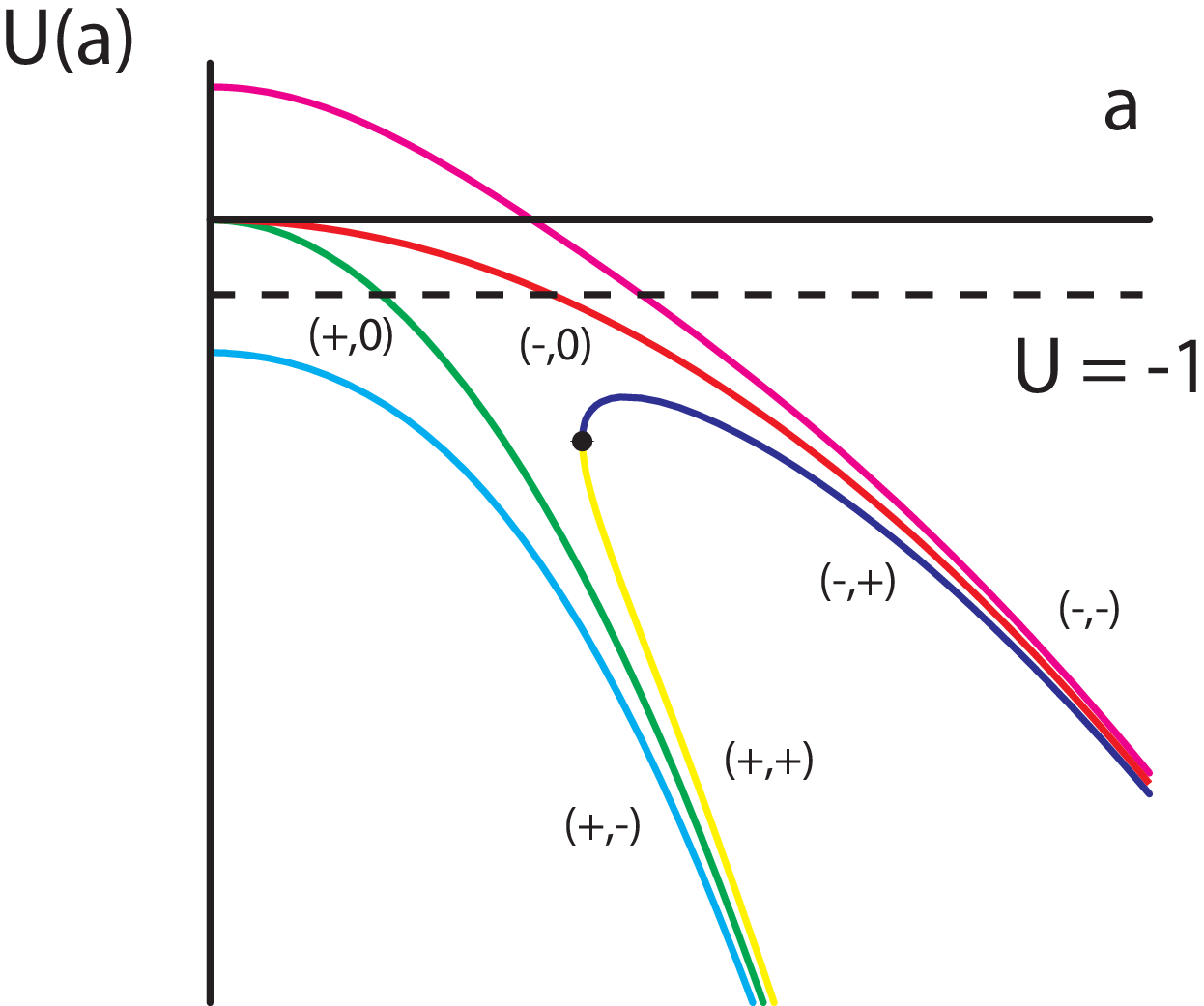}
\caption{The effective potential $U(\a,\eps,\mu)$ describing the
brane trajectories in various cases, classified by the
signs of $\epsilon,\mu$. The black dot indicates the singular radius $a_{crit}$
which demarcates the $\eps = \pm 1$ solutions with positive bulk mass,
and may be accessible by evolution depending on the
precise position of the line $U = -1$, which can shift
up and down as a function of the parameters. } \lab{fig:pot}}
Let us first reconsider the special case $\mu=0$ from this
perspective. In this case, the potential reduces to a simple
inverted harmonic potential,
\be U(\a,\eps,\mu=0)=-H_\pm^2\,\a^2\lab{simpU}\ee
where $H_-$ ($H_+$) corresponds to the choice $\epsilon=-1$ ($+1$).
Given that the effective energy is a fixed negative quantity, the
trajectories of interest approach the origin from infinity, bounce
off the potential at $\a_{min}=1/H_\pm$, defined in figure
\ref{fig:pot} as the point of intersection of $U(a)$ with the line
$U=-1$, and then head back to infinity. Of course, this is precisely
the behaviour shown by the analytic solutions (\reef{H}), describing
de Sitter spaces on the brane in global coordinates.

An alternative visualization of the DGP cosmology is given by plotting
the solution as a trajectory in phase space. Caution must be
exercised as the phase plot is {\it not} that
of an autonomous dynamical system, and hence can display irregularities,
nonetheless some qualitative features are very easily extracted. Explicitly,
to get a dimensionless plot, rescale variables:
\be
{\hat t} = H_0 t\;, \;\;\; {\hat a} = H_0 a \;, \;\;\;
{\hat \mu} = H_0^2 \mu \;,\;\;\; {\hat \sigma} = \sigma^*/H_0
\ee
and define the phase coordinates as
\be
X = \frac{1}{\hat a} \qquad Y = \frac{{\dot{\hat a}}}{\hat a}\ .
\ee
The Friedmann equation (and hence the trajectory in phase space)
coming from the square of (\reef{israel2}) now becomes
\be
(X^2 + Y^2)^2 - (1+ 2{\hat\sigma})(X^2+Y^2) + {\hat\sigma}^2
+ {\hat\mu} X^4 = 0
\ee
A set of typical plots (for the SA branch) is shown in figure \ref{fig:phase}.
It is easy to see that in the absence of a bulk black hole, the
trajectory is a circle in the $(X,Y)$ plane, whose radius is fixed by
the brane tension. We also see that for very small black hole masses,
the cosmology is very slightly perturbed, with the SA brane being
{\it repelled} by positive mass black holes.
Finally, we also notice that for large (positive) black hole masses, the
trajectories exhibit pathologies, which will be described in section \ref{psing}.
\EPSFIGURE[left]{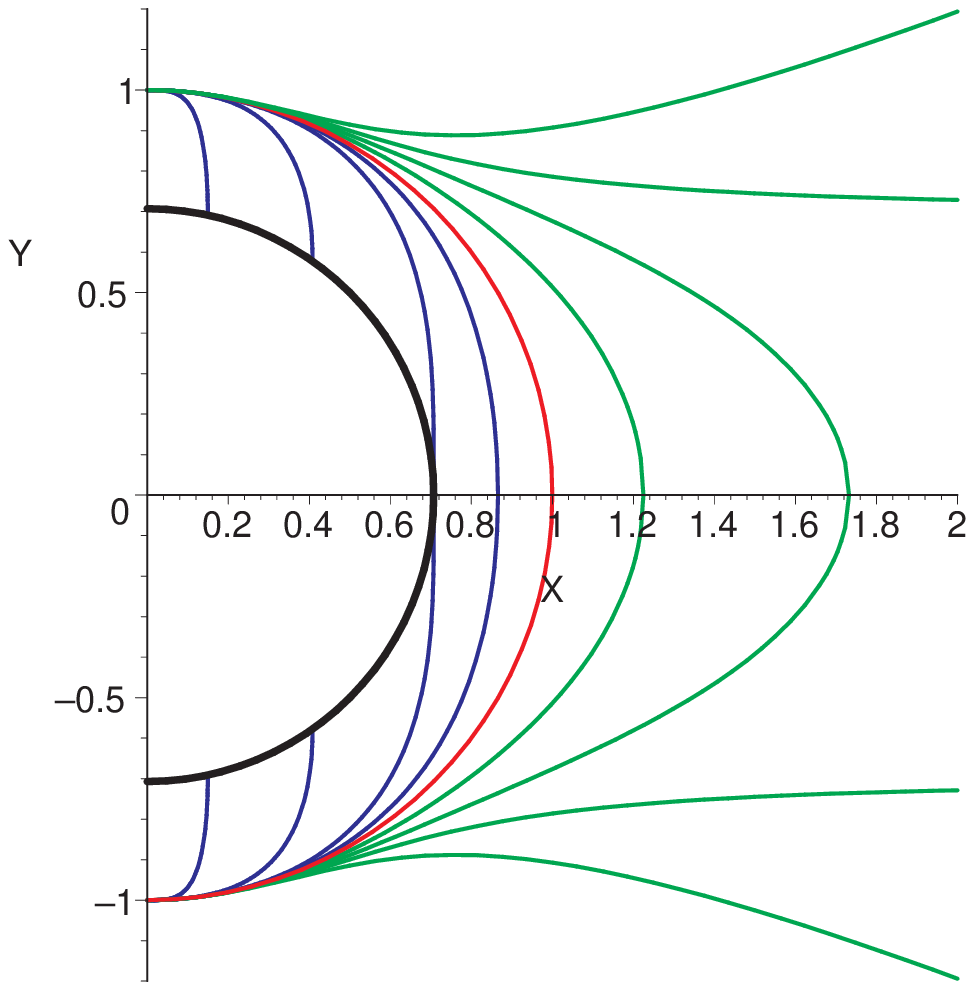}{A sample of phase plane
trajectories for the DGP cosmology with positive and negative bulk
black holes masses for the SA branch solution. We have taken
$\sigma=0$ for this plot.
The red circle (of radius one) represents
the ${\hat\mu}=0$ or standard DGP cosmology. The green plots are
with negative bulk black hole masses
(${\hat\mu}=-1/3,-2/3,-1,-3/2$), and the blue plots correspond to
positive bulk black hole masses (${\hat\mu}=1/3,1,9,500$). Note that
for ${\hat\mu}\ge1$, the trajectories terminate on the pressure
singularity (described in the text) which corresponds to the solid
black circle at $X^2+Y^2=1/2$.
Similarly trajectories with ${\hat\mu}<-1$ asymptote to
$(X,Y)\rightarrow(+\infty,\pm\infty)$, corresponding to reaching the
singularity at $a=0$.\lab{fig:phase}} 

It is interesting to note that if $\mu\rightarrow0$, we can regard
our $\mu \neq 0$ solutions as a small perturbation of the $\mu=0$
case. Hence in this limit, one should be able to relate our new
solutions to the fluctuation analysis \cite{CGKP,review}. We leave
the details of this analysis to appendix \ref{radion} but observe
that one finds that the perturbation introduced by small $\mu$ is
related to the homogeneous mode of the `radion'. In particular, on
the SA branch, this radion mode is normalizable and so describes a
dynamical field in the theory. Further, on the SA brane, since the
radion is always either a ghost or a tachyon, it is expected to be
associated with an instability. Indeed, in the appendix we will show
that at distances much greater than the gravitational radius $\mu$
of the bulk mass $m$, where we can trust linearized gravitational
fields in the bulk, the bulk mass parameter is completely encoded by
the nontrivial radion configuration. The mass is zero if the radion
vanishes, and its sign is determined by the sign of the radion. Now,
on the N branch, the radion is not a dynamical field in the theory,
as it is not normalizable. Of course, this reflects the fact that
our full solutions on the N branch contain a black hole (or a naked
singularity) at the center of the bulk space. However, describing
the latter requires the full nonlinear gravity theory no matter how
small $\mu$ is and so goes beyond a linearized analysis presented in
\cite{CGKP,review}. Thus on the normal branch $\mu$, and therefore
the bulk mass $m$, can be regarded as boundary conditions from the
point of view of the linearized theory. This does not exclude the
possibility that $\mu$ may be changed by processes whose initial
stages may be reliably described within the perturbative approach of
the linearized analysis, but the complete evolution necessarily goes
into a nonlinear regime. For example, one might consider a uniform
spherical wave emitted from the brane with a small amplitude and
then collapses at the center of the bulk space to form a black hole.

While the above comments are largely observations about mathematics,
we want to comment that the physics is distinguished here between
the SA and N branches. If we consider generic local processes which
result in transmitting energy off the brane into the bulk, for the
closed cosmology of the N branch, this energy simply disperses
within the finite spatial slices of the bulk geometry. In contrast,
on the SA branch, the analogous processes will generically send out
some energy which `leaks out' to infinity and so inevitably $\mu$
decreases. Thus, on the SA branch, and in general in the presence of
SA branes, even if we start with a bulk geometry where $\mu = 0$, we
will be able to access $\mu \ne 0$, and in particular $\mu < 0$.
Further, given the discussion above, we should be able to describe
this evolution within the linearized theory as radion dynamics. In
the bulk we can always go sufficiently far from the SA brane that
the linearized description will be valid, and so it is hard to see
how such processes can be avoided without a complete exclusion of SA
branes. On the other hand, if we ban naked singularities, such processes on
N-branch will not create configurations with $\mu < 0$.

\subsection{Positive black hole mass in the bulk and pressure singularities} \lab{psing}

We now consider the effect of a {\it positive} bulk mass ($\mu>0$).
As we have just discussed, this could be generated perturbatively by
the radion on the SA branch, or through a nonperturbative altering
of the boundary conditions on the N branch. Typical 
potentials (\reef{pot}) for this case are 
shown in figure \ref{fig:pot}. For any value of $\mu$, there are of course 
{\it two} branches for the potential, one  corresponding to the 
N branch ($\eps=-1$), and one corresponding to the 
SA branch ($\eps=+1$). When $\mu$ is positive, the two branches 
connect to one another at a singular point, indicated with the black dot.

Let us begin by looking at the SA branch ($\eps=+1$). At this point it 
useful to consider the phase plane plot given by figure \ref{fig:phase}. 
Here we are interested in  positive bulk mass, so we should turn our 
attention to the blue trajectories. These clearly split into two families: 
connected trajectories and disconnected trajectories. The  connected 
trajectories correspond to small values of $\mu$, and pass safely through 
the $Y$-axis. These trajectories are qualitatively similar to  the $\mu=0$ trajectory.  
A glance at the $(+, 0)$ curve for the potential (\ref{pot}) in figure \ref{fig:pot} 
indicates the following generic behaviour: the brane falls in from  large $a$ 
until it hits a bounce at some finite value $a_{b}$ and then expands 
back out again. The trajectory is completely non-singular and the bounce
occurs when the potential crosses $U=-1$. It follows that
\be
a_{b}= \frac{1}{\sigma_*}\left[   \half+\frac{\sigma_*}{H_0}
-\sqrt{\frac{1}{4}+\frac{\sigma_*}{H_0}-\sigma_*^2\mu}\right]^{1/2} \lab{ab}
\ee
Note that the brane always stays outside of the black hole horizon 
since $a_{b} \geq \mu^{1/2}$. It is easy to see that $a_{b}$ increases 
as we increase $\mu$, which agrees with our earlier  statement 
that the SA brane is  {\it repelled} away from positive mass black holes. 
The repulsive effect of $\mu > 0$ on the
SA branch is expected given the analysis of the effective Friedmann
equation (\reef{expand}) in which a negative density radiation term
appears with positive $\mu$ and $\epsilon=+1$. However, since we are keeping the
exterior or asymptotic region of the black hole geometry
(\reef{schwarz}) and, at the same time, the brane has a negative
effective tension, we cannot gain any futher intuition by
considering the global geometry.

The disconnected trajectories occur for larger values of 
$\mu$, greater than some critical value $\mu_{0}$, 
to be determined shortly. These trajectories terminate on 
the bold black circle in figure 3, %\ref{fig:phase}, 
at which point the brane runs into/out 
of a pressure singularity. This point also occurs precisely on the bold 
black dot on the $(\pm, +)$ curve in figure \ref{fig:pot}.  So what 
exactly is going on? Suppose the brane is falling in from a large $a$. 
At some point it will reach the black dot, at a critical value given by 
\beq 
\a_{c}=\left[\frac{\mu}{H_0^2/4+H_0\s^*}\right]^{1/4}\,. \lab{acrit}
\eeq
For $a<a_{c}$, the potential (\ref{pot}) is complex, so this regime is 
unphysical. As $a \to a_{c}$ from above, the potential terminates 
with an infinite slope at the black dot. At this point, $\dot a$ remains 
finite, but $\ddot a$ (and any higher derivatives) diverge.

We can confirm that this is a real physical singularity on the brane
by calculating the Ricci curvature for the induced metric, \eg
\be  \mathcal{R} =
6\left(\frac{\ddot\a}{a}+\left(\frac{\dot\a}{\a}\right)^2
+\frac{1}{\a^2}\right)\, ,
\lab{cinvar}
\ee
Clearly, this expression blows up at $\a = \a_{c}$,
since $\ddot\a$ diverges as $a \to
a_{c}$ and so we have a genuine singularity on
the brane.  This occurs at some finite radius which is
otherwise ordinary from the perspective of the bulk geometry. 
For a phenomenologically interesting range of parameters
(\eg the Schwarschild radius of the
black hole is smaller than the Hubble scale on the brane),
$\a_{c}$ is typically far outside of the bulk black hole
event horizon, as can be seen by comparing it with 
the bulk black hole horizon radius $\a_h=\mu^{1/2}$:
\be \frac{\a_h^4}{\a_{c}^4} = \mu\,\left(H_0^2/4+H_0\s^*\right)
= \frac{1}{12\pi^2}\,\frac{m}{M_5}\,
\left(\frac{M_5^4}{M_4^4}+\frac{\s}{6M_5^2M_4^2}\right)\,.
\lab{ratio}
\ee

This singular behaviour observed above resembles certain
`sudden' cosmological singularities studied recently \cite{sudden},
induced by the diverging pressures on certain spacelike
hypersurfaces as opposed to diverging energy densities. In fact,
these singularities are really reminiscent of the well known
pressure divergences in the McVittie solution \cite{mcvittie}, which
describes the field of a spherically symmetric mass in an FRW
universe. Here, we have the field of a black hole, and the matter
exterior to it is given by a thin brane. In the McVittie case also,
the singularity of the solution induced by the diverging pressure is
spacelike, and in the past extends well beyond the region where the
black hole horizon of the fixed mass would have been.

So, on the SA branch, when the black hole 
mass is small and positive ($0\leq \mu <\mu_0$), the brane 
follows a non-singular trajectory, falling in from large $a$, reaching 
a minimum at the bounce $a_b$, and expanding back out again. 
Here there is no pressure singularity because for these values of $\mu$ 
the brane hits the bounce before it can reach the point where the 
singularity would occur. In other words, $a_b>a_c$. For larger 
values of the black hole mass ($\mu>\mu_0$), the situation is 
reversed: a brane  falls in from large $a$ and hits a pressure singularity 
at $a_c$ before it gets the chance to bounce. This corresponds to 
the case $a_b<a_c$. Clearly we can calculate the critical 
value\footnote{The reader may note that equation (\ref{ab}) 
shows another critical value $\tilde \mu_0=(\sigma_*+H_0/4)/(H_0\sigma_*^2)$. 
However, $\mu_0<\tilde \mu_0$, and so the latter is not physically significant.}, 
$\mu_0$ by setting $a_b=a_c$, and solving for $\mu$,
\be \mu_0 = \frac{1}{H_0}
\frac{(\sigma^* + H_0/4)}{(\sigma^*+H_0/2)^2} \, . 
\ee

The behaviour on the N branch ($\eps=-1$) is fairly similar. Without 
going into details, we note that for small enough $\mu>0$, 
the behaviour is qualitatively the same as for $\mu=0$. Glancing at 
the curve $(-, 0)$ for the relevant potential in figure \ref{fig:pot}, we 
see that  the brane follows a non-singular trajectory,  falling into some 
minimum radius and bouncing out again. This time, the  position of the 
bounce {\it decreases} as $\mu$ increases. This means that a N brane 
is {\it attracted} towards a positive mass black hole.  Of course, 
this is exactly in accord with our intuition that
there is now a black hole at the center of the bulk geometry which
exerts a gravitational attraction on the brane. Similarly, the
appearance of an effective radiation term (with a positive density)
in the effective Friedmann equation (\reef{expand}) also reflects the
attractive effect of positive $\mu$ with $\epsilon=-1$.

At large enough values of $\mu$, the brane will once 
again run into a pressure singularity. This corresponds 
to following the curve $(-, +)$ in figure \ref{fig:pot}, hitting 
the singularity at the black dot. For completeness, we note 
that on the N branch, there is in fact a third scenario. This 
occurs for intermediate values of $\mu$, and corresponds 
to the case where the $(-, +)$ curve in figure \ref{fig:pot} 
crosses the line $U=-1$, but the black dot appears below 
the line. Then there are two possibilities: either we have a 
bouncing trajectory with no singularity (as for small $\mu$), 
or we have a brane that flows out of the pressure singularity, 
expands out to a maximum radius, and then contracts until it 
hits another pressure singularity.  In addition, tuning $\mu$ such that $U_{max}=-1$
also allows for special trajectories where the brane is static
and sits at a fixed radius corresponding to the maximum.

Although the pressure singularity occurs for large 
enough $\mu$ on {\it both} branches, the effects 
are actually very different, depending on which 
branch (N or SA) you are on. To see this, let us 
examine the nature of this pressure singularity more closely.
Intuitively, we expect that this singularity will be very dangerous
because the pressure divergence will yield to the perturbations of
the fluid dominating the universe becoming superluminal. Indeed, the
speed of sound of such perturbations is given by $c_S^2 =
\frac{\partial p}{\partial \rho}$, which will behave as $c^2_S \sim
p/\rho$ as $p \rightarrow \infty$. To check this, we can use a
relativistic field theory as a probe, and ask what happens to its
fluctuations as the universe approaches the singularity. Here we use
the formalism for studying small perturbations in closed FRW
universes that has been developed for inflation \cite{inflpert}. As
in a simple recent application \cite{shiu}, we start with $\nabla^2
\chi = 0$, expand the field $\chi$ in spherical harmonics $Y_{nlm}$
on $S^3$ as $\chi = \sum_{nlm} \chi(\eta)_n Y_{nlm}$, obeying $\vec
\nabla^2 Y_{nlm} = - (n^2-1) Y_{nlm}$ for $n-1 \ge l \ge |m|$ where
$\vec \nabla^2$ is the Laplacian on $S^3$, transform to the
conformal time coordinate $d\eta = d\tau/\a$, and define the new
field variable $\xi_n = \a \, \chi_n$. A straightforward calculation
yields the equation governing the time evolution of a particular
radial quantum number $n$,
\be
\xi_n'' + \Bigl(n^2 -1 - \frac{\a''}{\a} \Bigr) \, \xi_n = 0 \, ,
\label{perteq}
\ee
where the primes denote derivatives with respect to the conformal
time $\eta$. Now, it is easy to see that $\frac{\a''}{\a} = \a^2
(\frac{\ddot \a}{\a} + \frac{\dot \a^2}{\a^2})$ in terms of the
original time $\tau$. Further, differentiating equation
(\reef{classic}) with respect to $\tau$, we find $\ddot \a = -
\frac12 \frac{\partial U}{\partial \a}$.  

Now let us focus on what happens on the 
N branch. Consider the $(-, +)$ curve for 
the potential $U(a)$ in figure \ref{fig:pot}. 
As we approach the singularity (represented by the black dot), 
it is clear that the slope of potential is large and positive. It 
follows that as $a \to a_c$, $\frac{\a''}{\a}\to  -\infty$ , which 
converts equation (\reef{perteq}) into an harmonic
oscillator with a huge mass, for any value of the comoving momenta
$\propto n$. Therefore the singularity has the
effect of essentially decoupling the local dynamics of the field
theory inhabiting the universe. 

In contrast, on the SA branch, we can see from  looking at the 
$(+, +)$ curve in figure \ref{fig:pot} that the slope of the potential
becomes large and {\it negative} as we approach the singularity. 
It follows that as $a \to a_c$, $\frac{\a''}{\a}\to  +\infty$, indicating 
that a tremendous, exponential instability sets in
on all scales. Clearly, this applies to
the physical mode $\chi_n = \xi_n/a$ as well, because $\a$ remains
finite throughout. Of course, this instability could lead to
singularities when small perturbations are introduced to the
(homogenous) SA solutions on the bouncing trajectories where
$\a_b$ is not much bigger than $\a_{c}$, since
$\frac{a''}{a}$ still becomes very large near the bounce.

\subsection{Negative mass in the bulk}

We now turn our attention to the
case of a negative  bulk mass ($\mu<0$). Of course, typically we would
not consider the bulk geometry (\reef{schwarz}) with $\mu<0$ because
it contains a naked singularity at $r=0$. However, in our
construction of the SA branch solutions, we have excised
the bulk geometry inside the brane world-volume and
have kept only the exterior. Hence this
construction removes the singularity and produces a (potentially)
smooth solution for $\mu<0$ and $\epsilon=+1$. Furthermore, such a
solution can be generated perturbatively by exciting a normalizable
radion on the $\mu=0$ solution. 

We can learn more about the types of SA solution that occur for negative bulk mass by considering the green trajectories in the phase plane plot (Fig. \ref{fig:phase}). Again they are split into two families: connected trajectories for small $|\mu|$, and disconnected trajectories for large $|\mu|$. The connected trajectories correspond to the bounces, where the brane falls into some minimum radius and bounces back out again. As $|\mu|$ increases, the minimum radius actually {\it decreases}, so, as expected, we conclude that the negative mass has an {\it attractive} effect on the SA brane. The disconnected trajectories correspond to  $|\mu|>1/H_0^2$, when we  no longer have bounce solutions. The corresponding potential is given by the $(+, -)$ curve in figure \ref{fig:pot}. We can clearly see that  now the brane approaches
from infinity and `crashes' into the singularity in the bulk at
$a=0$ (or the time reverse). Since $a$ vanishes at this point (while
$\dot\a$ and $\ddot\a$ remain finite), a singularity also appears
from the brane perspective, \eg the Ricci curvature for the induced
metric \reef{cinvar} diverges. Unfortunately as it stands, the DGP
model is inadequate to determine the subsequent evolution of the
system. 
Some obvious, albeit naive possibilities are 1) that the
trajectory simply terminates at $a=0$ and a naked singularity
emerges in the bulk spacetime, or 2) that the trajectory might
continue on to negative values $\a$, which could be interpreted as
the brane passing through itself (and the singularity) at the origin
and emerging as an expanding brane with the opposite orientation. It
is noteworthy that at `early' times when the brane is still at
finite radius, the solution is smooth and so defines sensible
initial data.\footnote{Beginning from this smooth initial data, the
brane collapses to $a=0$ and the resulting singularity can be seen
by asymptotic observers. Hence cosmic censorship is violated in DGP
gravity. But perhaps this is not so surprising given that the DGP
branes do not satisfy any of the usual positive energy conditions
\cite{usual}, as observed above.} Hence even if the full evolution
seems to be singular, there is no obvious reason why these solutions
should be ruled out as physically unacceptable.

In contrast, on the N branch ($\epsilon=-1$) if we take $\mu < 0$, the N
branch construction will leave a naked singularity at the center of
the bulk space, which makes the solutions much less appealing. Nevertheless,
one could easily imagine that we are shielded from this singularity
by the presence of another SA brane in this geometry, evolving
concentrically with the N brane. The latter would, of
course, not effect the dynamics of the N brane, which will 
always correspond to a bounce. This can be seen by 
considering the $(-, -)$ curve in figure \ref{fig:pot}. As 
$|\mu|$ increases, so does the minimum radius, and so the effect of a
negative mass is repulsive on the N branch, as one might have
anticipated. 

All of this brings us to our main conclusion: with the SA branes, 
the DGP model admits physically sensible solutions for which the
five-dimensional energy can be arbitrarily negative. That is,
regarded as a five-dimensional theory of gravity and branes, the
spectrum of the DGP model is unbounded from below. Having identified
new negative energy configurations, it is natural to think that
these will be excited in both classical and quantum processes. While
finding explicit solutions which demonstrate the appearance of these
negative energy states in various dynamical processes is difficult,
it remains reasonable to assume that the theory should be
unstable.

\section{Bubbling SA branes}\lab{mix}

In considering possible instabilities and their description beyond
perturbation theory, given that the SA branes appear as the source
of the ills in the theory, it is natural to look at tunneling
processes in which the SA branes might be spontaneously created in
the Minkowski vacuum. In what follows we present a semiclassical
calculation for such a process using standard techniques of
Euclidean quantum gravity. With a straightforward application of
these techniques, we find that spontaneous nucleation of SA branes
in the bulk is unsuppressed. However, as we discuss below, these
calculations are not without certain ambiguities. We should also
note that the processes considered here are complementary to the
possible instabilities revealed in the previous section since the
calculations here are restricted to the case $\mu=0$. It would be
interesting to consider the interplay of these effects, which we
leave for another day.

We will now apply the standard instanton techniques for
the description of tunneling processes with Euclidean path integrals. The
tunneling probability from one configuration into another in an existing
Lorentzian geometry to leading order is
\be \m{P} \propto e^{-\Delta S/\hbar} \, , \lab{prob}\ee
where the instanton action $\Delta S$ is the difference of 
the Euclidean bounce and background actions,
\be \Delta S=S_\tt{bounce}-S_\tt{background}\, . \lab{delta}\ee
In analogy with particle mechanics,
we refer to the Euclidean solution describing the `bubble of SA
brane' at the center of five-dimensional flat space as the `bounce'.
The Euclidean action is given by
\be S_E=-2M_5^3\int_\textrm{bulk} \sqrt{g}
R-4M_5^3\int_\textrm{brane $+~\infty$} \sqrt{\gamma}
K-\int_\textrm{brane} \sqrt{\gamma}\left(M_4^2
\mathcal{R}-\sigma\right) \, , \lab{Eaction}\ee
where in the second integral, we have indicated the
integration of the extrinsic curvature over asymptotic infinity.
This is, of course, the standard Gibbons-Hawking term and in the
present case the net effect of the background subtraction will be
to cancel this term (which is otherwise divergent).

Upon Wick rotating to Euclidean signature, $t \to i t_E$, the bulk
space becomes simply $\real^5$
\begin{equation}
ds^2=dt_E^2+dr^2+r^2d\Omega^2_3\, , \lab{flat}
\end{equation}
with a ball of the radius $H_+$ removed and its surface identified with
the Euclidean geometry of the brane. Indeed, on the brane,
$\tau \to i \tau_E$ converts the de Sitter
geometry of the SA branch to that of a four-sphere with
\beq ds^2=d\tau_E^2 +\frac{1}{H^2_+}
\cos(H_+ \tau_E)^2 \, d \Omega_3^2\,.\lab{sphere}
\eeq
The instanton describing tunneling from empty space to a
space containing a single SA brane will consist of one half of both
of these geometries, but in squaring the wavefunction, the tunnelling
probability (\reef{prob}) requires the action for the full Euclidean
bounce. Note, that this construction is perfectly consistent with the $\mathbb{Z}_2$
orbifold condition. We can take two $5D$ Euclidean spaces with a ball removed, and
the remaining spaces identified across the spherical boundary, and then analytically continue
to Lorentzian signature at the equator, to get the SA brane. This will provide an `interior'
picture of the process, without any reference to the embedding space. Clearly,
we can also consider the processes where $\mathbb{Z}_2$ condition is lifted.

Now, to calculate the action for the bounce, we first use the
Euclidean version of the equations of motion
(\ref{bulk},\ref{israel}) to deduce that
\be R=0, \qquad 4M_5^3K+\frac{2}{3}M_4^2 \m{R}-\frac{4}{3}
\sigma=0\,, \lab{Eeom} \ee
where, as above, we have set all the matter contributions on the brane, except the tension,
to zero. Hence the Euclidean action
(\reef{Eaction}) for the instanton reduces to
\beq S_\tt{bounce}=-\frac{1}{3}\int_\tt{brane} \sqrt{\ga}
\left(M_4^2 \m{R}+\sigma\right)-4M_5^3\int_{\infty} \sqrt{\gamma} K \, .
\lab{Eact2}\eeq
Plugging in the spherical geometry (\reef{sphere}) above yields
\beqa S_\tt{bounce}&=&-\frac{1}{3H_+^3}\int
d\Omega_3\int^{\frac{\pi}{2H_+}}_{-\frac{\pi}{2H_+}}d\tau_E\,\cos(H_+
\tau_E)^3 \left(12M_4^2H_+^2+\sigma\right)-4M_5^3\int_{\infty}
\sqrt{\gamma} K
\nonumber\\
&=&
-\frac{8\pi^2}{9H_+^4}\left(12M_4^2H_+^2+\sigma\right)-4M_5^3\int_{\infty}
\sqrt{\gamma} K \, . \lab{Eact3}\eeqa
Now we must also consider the background contribution to the
instanton action (\reef{delta}). The background spacetime is simply
flat space and so the only contribution to the action is the
asymptotic integral of the extrinsic curvature, precisely
cancelling the second term
in (\reef{Eact3}). Thus the final result for (\reef{delta}) is
\beq \Delta S=
-\frac{8\pi^2}{9H_+^4}\left(12M_4^2H_+^2+\sigma\right)=
-\frac{8\pi^2}{3H_+^4}\left(8M_5^3H_++\sigma\right) \, . \lab{delta2} \eeq
We immediately note that that the sign of either of these expressions
is obviously {\it negative} for positive tension, $\sigma>0$. Hence this action
provides no suppression for the tunneling probability (\reef{prob})!

Our interpretation of this result is certainly not that the
tunneling probability is greater than one, but rather that the
saddle-point approximation implicit in (\reef{prob}) is not a good
one. A more careful analysis would be required to properly
evaluate the tunneling probability. However, what is clear is that
the breakdown of the saddle-point approximation implies that the
tunneling is not strongly suppressed and so there is large mixing
between the empty five-dimensional spacetime and that containing a
SA brane. By extension, one can infer that there is a large mixing
between all of the five-dimensional spaces containing any number of
SA branes. Hence we are led to conclude that empty five-dimensional
Minkowski space does not provide a good description of the quantum
vacuum of the five-dimensional theory. Note that while this conclusion
is in agreement with that found in the previous section, the present
calculations are restricted to the zero-energy sector and make no
reference to the negative energy states found previously.

Of course, it must be said that this rather dramatic conclusion
hinges crucially on the sign of the Euclidean action in
(\ref{delta2}). At the same time, there has been a long-standing
debate in the quantum cosmology literature
\cite{debate,harhaw,tunel} centered on precisely this sign. While we
have nothing new to add to that debate, we would point out that
there are significant differences between the setting of quantum
cosmology\footnote{The sign ambiguity reappears in the present
context when considering the N branch, in which the solutions
correspond to closed five-dimensional universes. If one evaluates
the Euclidean action (\reef{Eaction}) for a hemispherical brane, the
Hartle-Hawking wavefunction for this closed universe becomes
$\exp[-S_{HH}]$ with
\beq  S_{HH}=
-\frac{4\pi^2}{9H_-^4}\left(12M_4^2H_-^2+\sigma\right)=
-\frac{4\pi^2}{3H_-^4}\left(\sigma-8M_5^3H_-\right)\lab{HaHa}\,.\eeq
Note that the sign is again negative, as is standard for the
Hartle-Hawking wavefunction \cite{harhaw}, but alternate approaches
would flip this sign \cite{debate,tunel}. Note that taking the limit
$M_5^3\to0$ in the second expression above, reproduces the standard
result for four-dimensional Einstein gravity coupled to a
cosmological constant $\Lambda$: $S_{HH}=-2\pi^2\Lambda/3H^4$ where
$H=\sigma/6M_4^2$ and $\Lambda=\sigma/2$.} and the present
description of nucleating SA branes. Interpretational ambiguities
are naturally present in quantum cosmology where one considers the
quantum birth of a {\it whole} closed universe from `nothing'. In
contrast, in the setting describing the nucleation of SA branes, we
have a standard tunneling process describing the formation of a
`defect' in a larger empty spacetime, which serves as a {\it fixed}
background. Hence we can take it as the usual background describing
a metastable state, in which a tunneling process starts, and where the 
positivity of energy is defined by the ghost-free N-branch configurations. 
In this way we do not face the same interpretational issues as in the case of
`tunneling from nothing'. Although such interpretational issues are
absent, this still does not guarantee the sign of the action in the
tunneling amplitude. Indeed in a similar calculation,
ref.~\cite{vileni} advocated the opposite sign to that chosen in
(\ref{delta2}) -- however, we note that the discussion there
overlooked the boundary contribution of the Gibbons-Hawking term, 
as well as focussing on hypothetical brane embeddings with $K_{\mu\nu}=0$.

Let us consider then various choices for an unconventional analytic
continuation in the path integral. The first and simplest choice
would be choosing the opposite sign for the entire action, \ie
yielding $\m{P} \propto e^{-|\Delta S|}$. However, in this case, we
are implicitly applying the same unconventional continuation both to
the saddle-point with the spherical Euclidean brane and to that for
empty flat space. The latter seems rather problematic and so it is
doubtful that this can be the correct choice for the continuation.
As a compromise then, we could simply choose the conventional
continuation for empty space and choose the opposite sign for the
nontrivial saddle-point. However, in this case, the boundary terms
will not cancel between the two solutions and so resulting $\Delta
S$ is divergent. Hence this approach is also problematic. The final
alternative which we consider here is that the unconventional
continuation would only be applied to the brane action, \ie we would
only reverse the sign of the brane contributions in (\ref{Eaction}).
This choice may seem natural since the SA branes have already been
identified as the source of problems in DGP gravity and further this
continuation leads to a finite positive $\Delta S$. However, in this
case, the analysis above must be reconsidered. In particular, the
sign of the brane terms in the Euclidean equations of motion are
also reversed and so the tunneling instanton will contain a
Euclidean brane of radius $1/H_-$, rather than $1/H_+$ as above --
recall $H_\pm$ are defined in (\reef{H}). Hence, this approach does
not seem to produce a consistent description of the tunneling event,
in which the Euclidean brane naturally connects to the SA brane in
Lorentzian signature. Hence, while we can not at present guarantee
that the tunneling calculation does not require an unconventional
analytic continuation, we would say that straightforward calculation
of the Euclidean action seems the most appropriate choice.

It would be interesting to address this question using Hamiltonian
techniques, such as in \cite{fisch}. In a problem with some elements
in common with the present case, this approach was recently used to
argue the -- perhaps -- surprising result that the nucleation of domain
walls comprised of ghost matter should be suppressed \cite{ghosto}.
But we would also note that at least in certain situations, applying
such Hamiltonian techniques to quantum gravity is known to produce
erroneous results \cite{hfire}.

We must comment, however, that the conventional analytic
continuation for the tunneling calculation is not free of further
technical complications, as we now describe: The Euclidean instanton
calculation above could possibly describe a number of different
processes: {\it i)} decay of five-dimensional Minkowski space by the
spontaneous creation of a SA brane; {\it ii)} the decay of a SA
brane into empty five-dimensional space; or {\it iii)} a mixing
between two distinguished components of the wave-function over
five-dimensional geometries, in the sense of quantum gravity. To
distinguish between these different possibilities in the
saddle-point approximation, one may evaluate the fluctuation
determinant around the tunneling solution and determine the number
of negative eigenvalues. This computation is complicated in the
present situation both by the breakdown of the saddle-point
approximation and by the fact that the Lorentzian analysis of the
fluctuations around the SA brane indicates the presence of a
negative energy ghost. Hence we may expect an infinite number of
negative eigenmodes for the bounce solutions. A more careful
analysis would be required to determine which of the processes
listed above is really relevant. Above, we have simply referred to
the process as `mixing' or `tunneling' in some general sense. In any
event, we expect this quantum tunneling to strongly modify the
classical physics, described by the simple solutions discussed, \eg
in section \ref{model}.
%nk
Note, however, that this result may in fact be an indication of the
presence of ghosts beyond linear perturbation theory. Indeed, we may
follow the converse argument, which simply tells us that since the
saddle point approximation breaks down, there should be unstable
directions in the phase space of the system. This would then suggest
that the strong coupling effects, argued to be important in
perturbation theory \cite{dispute}, may not remove the ghost on
their own, as long as they leave the self-accelerating solutions in
the spectrum.

Another interesting complication arises with the `wormhole'
interpretation of the SA branes mentioned above. Here some further
refinement is required in specifying the precise tunneling geometry.
Again, the conventional configuration implicitly has the SA brane
connecting two disjoint Minkowski spaces. Given two parallel
Minkowski spaces, instantons could be constructed in which a SA
brane appears in each of the two spaces, but there are two other
instantons for which both sides of the SA brane appear in either one
of the Minkowski spaces. Hence quantum tunneling seems to naturally
lead us to think about configurations where the SA brane connects
distant locations within the same five-dimensional space and so
makes their role as a `wormhole' evident. The path integral leading
to the probability (\reef{prob}) would also include an integral over
the position of the instanton. In a conventional setting this would
produce a factor of the volume of the five-dimensional space. This
result is regulated by dividing out this factor to yield a
probability per unit time and per unit spatial volume. However, in
the situation where the instanton creates a Lorentzian wormhole,
there would be in fact an integral over the position of both sides
of the wormhole yielding two volume factors. In this case then, the
result still diverges if one asks for the probability that a SA
brane appear at given position and time in the five-dimensional
space. Thus the interpretation of these instantons appears less
straightforward than usual. Of course, both of these complications
are again eliminated if the branes are defined to be
$\mathbb{Z}_2$-orbifolds, in which case there is only one side to
the branes.

As a consequence of this tunneling processes (irrespective of the
sign issues above), we are naturally led to consider a
five-dimensional space with a single asymptotic region that contains
more than one SA brane (or a single wormhole). At some point in the
future evolution of these solutions, the branes will collide with
each other. Similarly, SA branes created inside a N brane universe
will typically lead to a collision between the two branes, \ie the
self-accelerating brane `crashing' into the normal brane. Clearly,
unless the tunnelling rates found above can be suppressed in some
way, this issue remains secondary, since the sheer proliferation of
the SA branes, which gobble up space like Witten's `bubbles of
nothing' \cite{witten} (even though there may be no `nothing' here
when we impose $\mathbb{Z}_2$ symmetry), will end up completely
destroying the bulk environment.

Above we considered tunneling processes in which a SA brane
is nucleated in empty five-dimensional Minkowski space. This
process would allow the quantum creation of such a brane in any
region of Minkowki space. In particular, consider the N-branch
where the brane encloses a region of Minkowski space with finite
spatial extent, as described in section \ref{model}. The minimum
radius of this region (at $t=0$) is $H_-^{-1}$, which also
corresponds to the size of the corresponding Euclidean solution.
Similarly the minimum radius of the SA brane (or maximum radius for 
the Euclidean bounce solution described above) is $H_+^{-1}\,<\,H_-^{-1}$.  
Hence it is easy to fit a SA brane worldvolume inside that of a N brane, for 
a fixed value of the tension, $\sigma$. Given the latter fact, it is 
straightforward to extend the above calculations for the spontaneous 
appearance of SA branes in the bulk space enclosed by a N brane. 
In fact, $\Delta S$ is precisely the same as given in (\reef{delta2}) 
and so the conclusions are the same for this case.

To close this section, we list some more exotic possibilities for
tunneling processes. First off, for $\mu < 0$ there is a very
interesting possibility for mixing between unregulated bulks with
naked singularities and their regulated variants involving SA
branes. The brane trajectories will come from a Wick-rotated version
of (\reef{classic}) and will have a smooth intrinsic geometry, even
though they close off at the singularity. However, near the
singularity their extrinsic curvature must diverge. If we were to
ignore the action on the singularity we would expect to find a
finite negative action. Clearly, this answer would be sensitive to
higher order corrections and hence unreliable. Yet, we would expect
in general that solutions with large negative actions should exist,
again yielding strong mixing, and allowing for a formation of a
naked singularity. Note, that the usual positive energy theorems
\cite{usual}, which normally enforce cosmic censorship, do not apply
in the DGP model. 

For the case $\mu>0$ we would expect that
nonperturbative processes describing bulk black hole formation are
possible on the N branch. These however may be suppressed by the
thermodynamic factors which do apply on the normal branch, in the
bulk. Another possibility would involve transitions between
collapsing solutions that would bounce to those which crunch, and
back. Without a precise calculation of the solutions it is not clear
what the significance of this instanton is. Indeed, for sufficiently
small $\mu>0$ there are no crunching, confined trajectories to begin
with.

\section{Discussion}\lab{discuss}

The DGP model originally captured the attention of theoretical
physicists by offering an explanation of the cosmic
acceleration that was distinct from the standard one with a
cosmological constant. An important element of the model was seen to
be its internal consistency. However, recently flaws in this
consistency have emerged. A detailed analysis of the
perturbative fluctuations revealed that the spectrum of the
self-accelerating solutions contain ghost-like and tachyonic
excitations \cite{CGKP}. In the present paper, we have taken a new
perspective on the model and rather than focus on the
four-dimensional physics of the branes, we have studied the theory
from a five-dimensional point of view. With this new perspective,
simple calculations straightforwardly led us to see a number of new
pathologies of the DGP model.

One simple observation in section \ref{model} was that the
conventional SA brane connecting two asymptotically flat regions can
be regarded a `wormhole'. As such, DGP gravity faces certain
problems which are inherent to such wormhole configurations
\cite{worm}. One issue that is often discussed in this context is
the necessity for negative energy densities to support a wormhole.
As was also discussed in section \ref{model}, the effective stress
tensor (\reef{stress}) of the DGP branes does not satisfy any
positive energy conditions and in fact, the SA brane behaves as a
brane with a negative effective tension. Another issue that arises
with wormholes is the possible appearance of closed time-like curves
(CTC's). Here one must note that the present wormholes differ from
those conventionally discussed \cite{worm} since the size of the
wormhole throat is not fixed, \ie the brane follows a hyperbolic
trajectory in Minkowski space. However, it is straightforward to
show that this dynamic behaviour does not prevent the appearance of
CTC's. In section \ref{mix}, we also found that the wormhole
geometries lead to additional divergences in the semiclassical
tunneling calculations. Of course, all of these problems are
eliminated if the DGP branes are $\mathbb{Z}_2$-orbifolds, in which
case, the brane is actually a boundary of the five-dimensional
spacetime.

In section \ref{negative}, we provided a construction of SA
solutions where the five-dimensional energy was arbitrarily
negative. While these solutions are singular at $a=0$ for a sufficiently
large and negative energy, they still admit well-behaved smooth
initial data and so there do not appear to be any obvious reasons not to include
these solutions. The appearance of a singularity is simply a
shortcoming of the model and a reminder that we are working with an
low energy effective theory. Hence our conclusion is that regarded as
a five-dimensional theory of gravity and branes, the spectrum of the
DGP model is unbounded from below. This reinforces the
expectation developed from the fluctuation analysis \cite{CGKP} that
the SA brane should be unstable to decay by both classical and
quantum processes. In particular, we have found explicitly (see the Appendix)
that the negative mass configurations can be smoothly attained by
classical radion variation from SA branes. Hence one can regard 
the solutions in section \ref{negative} as nonperturbative excitation of 
the normalizable radion mode, going beyond the original linearized analysis 
and extending the associated instability to the fully nonlinear regime of the theory.

As our construction in section \ref{negative} shows the addition of
DGP branes to the five-dimensional Einstein gravity has allowed the
theory to resolve the singularity of negative mass Schwarzschild
solution. From the point of view of a quantum gravity, this was long
ago argued to be a disaster for the theory \cite{prize}, for
essentially the same reasons as elaborated above. Namely, the complete
theory would admit negative energy states and the empty space would
no longer be a stable vacuum.
One might wonder if there might be some
super-selection rules that keep the energy positive. After all, for desired
phenomenological application, one needs SA branes which are really big, even with negative energy,
and so might be hard to produce. But the tunneling calculation, and classical radion
evolution seem to support that the opposite is true, and that the system will probe arbitraily
low energies.

While the explicit constructions of section \ref{negative} are very
symmetric, it is clear that they only represent a small corner of a
vast space of negative energy solutions. The key point is that in
the approach elucidated in section \ref{model}, the SA brane excises
a finite region at the center of the five-dimensional bulk and so
one can allow any manner of singularities there. A simple extension
of our symmetric solutions would be to choose the bulk geometry as a
negative mass Schwarzschild but to position the singularity away
from the center of the space and closer to the minimum radius
reached by the brane. Determining the precise trajectory of the SA
brane would be computationally a more complicated task but it seems
clear that the result will be a negative energy solution where,
rather than being uniformly spread, the excitation is localized in
one region of the brane. Ergo, inhomogeneities with arbitrarily negative energies
should also form. Based on this and the
connection to the perturbative analysis, one should expect that to encode such
modes there should be nonlinear solutions which represent excitations of
the higher unstable modes as well, \ie ghost or tachyon modes with
nontrivial angular momentum on the three-sphere.

It is interesting to contrast our present results on negative energy
states with previous observations of effective negative energies
associated with matter configurations on SA branes
\cite{point,wall}. In both cases, these nonlinear solutions showed
behaviour corresponding to a negative energy density at some
intermediate scale. However, this effect is not indicative of the
overall mass or energy measured at asymptotic distance scales. In
the case of the domain walls \cite{wall} -- see also \cite{kyony} --
where the full five-dimensional solution is known, one clearly sees
that the $5D$ wormhole geometry of the SA brane screens the
effective negative energy. That is, since the asymptotic geometry is
simply $5D$ Minkowski space, from the five-dimensional perspective
advocated here, we see that these configurations correspond to
precisely zero energy states. It is still interesting to ask if
these effective negative energies appearing at intermediate scales
can produce instabilities through some local processes on the brane.
However, we would argue that this is unlikely to be the case. By
considering domain walls connecting regions with different brane
tensions, it is clear that the negative energy is screened on the
scale set by the size of the bubble surrounded by the domain
wall.\footnote{The solutions given in \cite{wall} describe bubbles
of the size of the cosmological horizon. By considering the
analogous constructions with domain walls separating regions with
different brane tensions, we find solutions where the size of the
bubble is disentangled from the horizon scale.} Hence it seems that
this screening will prevent the formation of such bubbles in
dynamical processes.

In section \ref{negative}, we also saw how a {\it positive} energy
bulk would lead to problems of a very different nature. Again, as
shown in the appendix, one can easily generate a positive mass in
the bulk by exciting the radion on the SA background. Typically, the
cosmological solutions we found run into a pressure singularity on
the brane, at some finite value of the scale factor. This represents
an absolutely disastrous instability on the SA branes, as can be seen by considering
the evolution of a scalar field on the brane. As the pressure
diverges, so the scalar field starts to behave as if it has a
diverging tachyonic mass on all scales! The pressure singularity can
also appear for a positive energy bulk on the N branch, but in that
case the scalar field picks up a diverging positive mass, and simply
decouples from the dynamics of the universe.

Yet another pathology came from standard semi-classical calculations
showing that there should be a rapid spontaneous creation of SA
branes in any region of five-dimensional Minkowski space. Of course,
as discussed in the previous section, these calculations are not
free of various ambiguities inherent in the Euclidean path integral
in quantum gravity. However, having alerted the reader to these
caveats, we continue to consider the implications of the results.
Note that these explicit calculations are restricted to the
zero-energy sector of the five-dimensional theory. Hence this
instability is complementary to the instabilities which one would
expect to arise from negative energies discussed above. In any
event, these results suggest that empty five-dimensional Minkowski
space is not the vacuum and not even close to the vacuum of the DGP
model as a five-dimensional theory of gravity. Of course, this is a
disturbing result that calls into question any intuition or results
derived from classical solutions in the DGP theory. In particular,
it casts serious doubt on the SA solutions, which are really the
reason behind these pathologies. Interestingly, this is reminiscent
of the situation encountered in higher derivative gravities in $4D$,
where classical de Sitter solutions supported by higher derivative
terms were also linked to vacuum instabilities \cite{simon}. There,
a cure which was devised was to exclude such configurations from the
allowed set of solutions of the theory.

Hence what are we to conclude from these results? In fact, the only
firm conclusion we wish to draw is that as a theory of
five-dimensional gravity the DGP model is not consistent, as it
displays some severe pathologies. We must remind ourselves, however,
that this model is only a low-energy effective model and not a
complete theory of gravity or other physics. As we noted above, we
are readily reminded of this status by the appearance of spacetime
singularities from rather generic and uneventful initial data. Could
it be then that the DGP model has a UV completion, which evades all
of the problems which we have elucidated here? While this may seem a
rather unlikely eventuality, let us point out a few simple examples
where in fact this might be the case.

The first example is provided by \cite{donmark}, which was motivated
by the appearance of negative-tension branes in the original
Randall-Sundrum model \cite{RS}. Of course, the latter have some
obvious instabilities that were eliminated by invoking a
$\mathbb{Z}_2$-orbifold boundary condition. The authors of
\cite{donmark} found a new dynamical instability, which applied even
with this boundary condition, by studying black holes falling in on
negative tension branes from the bulk. This process was found to
produce a catastrophic `big crunch' singularity which caused the
entire bulk space to collapse. However, they also observed that, if
the negative tension brane in the RS model is actually realized by
an orientifold of a higher dimensional geometry in a string theory
construction, then this instability should not be expected to arise.
That is, this pathology of the RS model is only a pathology of the
low energy description and so would not play a role in a string theory
realization of the same physics.

As a second example, we address directly one of our DGP pathologies,
albeit not with a UV completion of the model. Imagine
extending (\reef{action}) by adding the following Gauss-Bonnet 
%nk proposed change
term\footnote{Gauss-Bonnet terms in the bulk have been widely explored in
various braneworls setups \cite{gaussbonnet}. Here we only add them to the
brane action, and not in the bulk. Thus they are purely topological, being relevant
only for the quantum dynamics of the theory.}
%nk end change
to the brane action:
\beq \frac{\beta}{128\pi^2}\,\int_\textrm{brane}
\sqrt{-\gamma}\left(\mathcal{R}_{\mu\nu\rho\sigma}
\mathcal{R}^{\mu\nu\rho\sigma} -4\mathcal{R}_\mn \mathcal{R}^\mn
+\mathcal{R}^2\right)\lab{gb}\eeq
where $\beta$ is a dimensionless constant. In fact, the effect below
could be accomplished by adding any term involving squares of the
intrinsic curvature. A generic term would, however, modify the
behaviour of gravity on the brane at short (and long) distances and
also lead to new ghost excitations in the UV. Both of these
complications could be evaded with the Gauss-Bonnet term (\reef{gb}), which
because of its topological nature does not effect the dynamics.
However, if this term is carried through the calculations of section
\ref{mix}, the final result would be that $\B$ is shifted by a
constant, \ie (\reef{delta2}) is replaced by
\beq \Delta S'=
-\frac{8\pi^2}{9H_+^4}\left(12M_4^2H_+^2+\sigma\right)+\beta\,.
\lab{delta7}\eeq
Since the standard contribution is determined by the microscopic
parameters, $M_5$, $M_4$ and $\sigma$, once these are fixed to
produce a certain dynamics, we have the freedom to tune $\beta$ to a
sufficiently large positive value so that $\B'>0$. As a result, the
rapid spontaneous creation of SA branes would be suppressed and so
this pathology would be removed. While it may seem unnatural that
$\beta$ should have to be tuned in this way rather than simply being
$O(1)$ (or perhaps $O(128\pi^2)$ given our normalization), we would
point out that the DGP model already required a certain amount of
fine-tuning, \eg in the ratio $M_5/M_4$ to produce a
phenomenologically viable model. So while we have not addressed the
question of a UV completion, we have alluded to a theoretical
mechanism which may suppress\footnote{Recall that any tunneling at
all gives rise to configurations where SA branes will collide.} the
tunneling processes\footnote{Having pointed
to a positive influence of a Gauss-Bonnet term, we should also
reiterate the concern about the theory where the regularity of the
solutions found from analyzing leading order terms in the derivative
expansion is so sensitive to the inclusion of higher-dimension
operators, on the brane or in the bulk. This, in the very least,
underlies the importance of designing a UV completion before jumping
to phenomenological applications of the solutions of the theory at
the present stage.}  considered in section \ref{mix}. 
We should also mention, that in principle one
might also seek to suppress the nucleation of SA branes, or exclude
this channel altogether, in a similar vain as the Witten's bubbles
of nothing are suppressed in the usual KK compactification. This can
be done by adding other bulk degrees of freedom, for example
fermions, that will twist nontrivially in a bulk with an SA brane,
by seeking boundary conditions sensitive to the extrinsic curvature.
Such terms could, at least in principle, change the action of the
Euclidean SA configuration. It would be interesting to explore this
in more detail.

Our discussion then brings to the fore the importance of finding a
proper UV completion for the DGP model. While there are claims that
DGP gravity can be realized in certain string theory constructions
\cite{elias,ignatios} -- see also \cite{lowe} -- more recent
analyses yielded the argument that in fact it would be impossible to
find a UV completion for this model \cite{noUV}.\footnote{The
applicability of this analysis to DGP gravity was subsequently
called into question in \cite{nonon}.} Clearly it is very important
to resolve this issue and in fact, at present, this seems to be the
most pressing question to be addressed with respect to DGP gravity.
We contrast this situation with that of, \eg Randall-Sundrum
`gravity' \cite{RS}. The RS model is known to suffer from
gravitational instabilities involving unstabilized negative tension
branes. However, the basic five-dimensional model still serves as a
useful testing ground for new ideas in particle phenomenology (for
reviews see \cite{rubcsaki}), which stay away from the complications
of gravitational nonlinearities. Moreover, with the development of
the stabilizing mechanisms \cite{goldwise}, the instabilities can be
put under control at low energies. Further in the RS case, we can
rest assured that string theory provides consistent constructions
\cite{throat} which produce very similar physics to the simple
five-dimensional models and which evade the problems found in the
low energy effective theory. On the other hand, the DGP model was
constructed precisely as a testing ground for new gravitational
physics and although our discussions address nonlinear
instabilities, we know that these instabilities link up with the
ghost problems found in the linear theory \cite{CGKP}. Hence one
must object that in its present form the DGP model does not provide
a consistent framework to test new gravitational ideas, as long as
it permits `unchecked' SA branes. Thus we re-iterate that the most
pressing question to resolve about DGP gravity is whether a UV
completion exists, and, if so, what are the UV characteristics of
the theory. For example, perhaps a sensible UV completion can guide
us to make some simple modifications of the original DGP model which
will ameliorate the problems discussed above. In any event, the
present discussion reinforces that it is premature to use the DGP
model to develop any detailed cosmological tests.

Going beyond the DGP model, we note that the perspective and
analysis that we applied here can be utilized broadly in many other
braneworld extensions of four-dimensional Einstein gravity. Examples
which combine DGP and RS features were studied in, \eg
\cite{CGP,joel}. Our approach would give a quick diagnostic of the
consistency of such models. One expects that in the realizations
which contain perturbative ghosts and SA branches, similar negative
energy problems and singular evolutions will arise. It would be
interesting to check this explicitly. Also, it would be interesting
to explore higher-codimension models with brane localized terms.
Such models have been recently investigated in \cite{kaloper2},
where a construction without ghosts was provided. Yet it appears
that for some boundary conditions on the brane a nonlinear analysis
as in section \ref{negative} may yield negative energy solutions.
Finding out the precise link of these configurations could shed more
light on the interconnectednes of perturbative ghosts and possible
instabilities. Ultimately, learning more about such problems may
give us an opportunity to test the robustness of General Relativity.
While seeking modified gravities is an interesting endavour which
may be motivated by various cosmological and phenomenological
problems, it's also a dangerous venture. Without a reasonable UV
completion to guide us, many of these models will simply provide
pathological theories that do not have well behaved low energy
limits, and remain altogether unreliable. Yet, at this time we still
do not know for a fact if the successes of General Relativity really
require General Relativity, or might be reproduced by a more exotic
structure. Searching for more clues to point us either way thus
remains an interesting effort.

\acknowledgments We thank Christos Charmousis, Ed Copeland, Roberto
Emparan, Steve Giddings, Gary Horowitz, Alberto Iglesias, David E.
Kaplan, Andrei Linde,  Gustavo Niz, Minjoon Park, Oriol Pujolas,
Simon Ross, Paul Saffin, and especially Gia Dvali and Don Marolf for
helpful discussions and comments. We would also like to thank Ross
Stanley for his contributions towards the phase plane analysis. RCM
and NK thank Tokyo Institute of Technology for kind hospitality in
the beginning of this work. AP thanks UC Davis HEFTI program for
support and hospitality. Research at the Perimeter Institute is
supported in part by the Government of Canada through NSERC and by
the Province of Ontario through MRI. RCM acknowledges support from
an NSERC Discovery grant and from the Canadian Institute for
Advanced Research. Research at the KITP was supported in part by the
NSF under Grant No.~PHY05-51164. The research of NK was supported in
part by the DOE Grant DE-FG03-91ER40674 and in part by a Research
Innovation Award from the Research Corporation.

\appendix

\section{The radion and the Schwarzschild bulk}\lab{radion}

In section~\ref{negative} we examined in some detail the behavior
of a brane embedded in a bulk Schwarzschild geometry. For very small
values of the black hole mass, we observed that the bulk geometry is
only slightly deformed from flat space\footnote{As noted below, this
only really applies on the SA branch.} (\reef{minkowski}) and the
brane trajectory is only slightly perturbed from the standard
cosmology (\reef{cosmosol}). Hence in this regime, we should be able
to connect the new solutions of section \ref{negative} with the
fluctuation analysis of \cite{CGKP,review}. We commented above and
will demonstrate here with a detailed calculation that the small
$\mu$ perturbation corresponds to the homogeneous mode of the
radion, as studied in~\cite{CGKP,review}. Further, we observe that
on the self-accelerating branch, these perturbations are
normalizable, whereas on the normal branch they are not. Perhaps
this is not surprising since the normalizable radion decouples on
the normal branch, but not on the SA branch.

Before embarking on the detailed calculation, let us note that the
radion field roughly measures fluctuations in the brane's position
in the bulk. In going from a Minkowski bulk to a Schwarzschild bulk,
we would expect the brane to respond by being drawn towards the
black hole, and so clearly the radion field will be excited first
and foremost. Secondly, on the self accelerating branch
the black hole has itself been cut out, only leaving its long range fields
in the asymptotic region. This means that the boundary conditions `deep
inside the bulk' correspond to boundary conditions in asymptotically
flat space, \ie they are identical to those for a five-dimensional
Minkowski bulk geometry. Therefore, the Schwarzschild solution must
correspond to a {\it normalizable} perturbation. Of course, this is
not so on the normal branch, which retains the interior region.
Hence in that case the boundary conditions deep inside the bulk
differ: for the Schwarzschild bulk, we have a black hole while with
a Minkowski bulk, we simply have empty flat space. Therefore, the
Schwarzschild solution on the normal  branch
must correspond to a {\it non-normalizable}
perturbation, which simply means that this mode enters the nonlinear
regime deep in the bulk.

We will now elaborate our claims with an explicit calculation. When the
mass parameter, $\mu$, is small, the bulk solution (\reef{schwarz}),
and the brane equation of motion (\reef{israel2}) clearly correspond
to vacuum perturbations about the background solutions with a
Minkowski bulk ($\mu=0$). Linearizing in $\mu$, the bulk solution
(\reef{schwarz}) becomes
\be ds^2=d \bar
s^2+\frac{\mu}{r^2}(dt^2+dr^2)+\mathcal{O}(\mu^2)\lab{schpert}\ee
where $d \bar s^2$ is the background Minkowski metric
(\reef{minkowski}), written in global coordinates. We now change to
a de Sitter slicing of Minkowski by introducing new coordinates as
follows
\be r=\frac{e^{\epsilon Hy}}{H} \cosh (H \tau)\,, \qquad
t=\frac{e^{\epsilon Hy}}{H} \sinh (H \tau)\,. \ee
The background bulk metric now takes the form given in~\cite{CGKP,
review}
\be d \bar s^2=e^{2\eps Hy}\left[ dy^2+\bar \gamma_{\mu\nu}
dx^\mu dx^\nu \right], \ee
where $\bar \gamma_{\mu\nu}$ is a spherical slicing of $4D$ de
Sitter (see  equations (\reef{cosmogeo}) and (\reef{cosmosol})), and
$\epsilon=\pm 1$ depending on whether we are on the N branch
($\epsilon=-1$) or the SA branch ($\epsilon=+1$).

Using (\reef{schpert}), the Schwarzschild geometry now corresponds to
a perturbation of the form \be\delta g_{yy}=\delta g_{\tau \tau}=\mu
H^2 (1+\tanh^2(H \tau)), \qquad\delta g_{y \tau}=2 \mu H^2
\tanh(H\tau), \ee and $\delta g_{ab}=0$ otherwise. In \cite{CGKP,
review}, we predominantly worked in Gaussian-Normal (GN) gauge, for
which, $\delta g _{ay}=0$. Since we wish to identify the
Schwarzschild bulk with vacuum perturbations appearing in
\cite{CGKP, review}, it is therefore convenient to transform to GN
gauge as follows
\be
y \to y+\eta, \qquad x^\mu \to x^\mu +\xi^\mu
\ee
where
\begin{eqnarray}
\eta &=&-\frac{\mu}{2}\eps H e^{-2\eps Hy}\left[1+\tanh^2(H \tau) \right]
\lab{bender} \\
\xi^\mu &=& \frac{\mu}{4}e^{-2\eps H y} D^\mu \left[\tanh^2(H\tau)+4
\ln (\cosh(H\tau))\right]
\end{eqnarray}
Here $D_\mu$ is the covariant derivative for the $4D$ de Sitter
metric, $\bar \ga_\mn$. The metric perturbation now takes the form
given in \cite{CGKP, review}
\be
\delta g_{ay}=0, \qquad \delta g_\mn=e^{\eps Hy/2} h_\mn(x, y) \lab{GNsoln}
\ee
where
\be h_\mn(x, y)=\left[ D_\mu D_\nu +H^2 \bar \gamma_\mn\right]
\phi(x, y), \qquad \phi(x, y)=e^{-\eps Hy/2}\hat \phi(\tau), \ee
\be \hat \phi(\tau)=-\mu H \sinh(H\tau) \int^\tau \frac{d
\lambda}{\cosh^3(H \lambda)\sinh^2(H \lambda)}\lab{xxxx}\ee
It is easy to check that $(D^2+4H^2)\hat \phi(\tau)=0$, which means
that $h_\mn(x, y)$ is transverse-trace-free. Indeed, if we compare
this with the perturbations given in \cite{CGKP,review}, we see that
the mode $\phi(x, y)$ can indeed be identified as the radion. On the
SA-branch ($\eps=+1$), the mode decays for large $y$, and is
therefore normalizable. In contrast, on the N branch ($\eps=-1$),
the mode grows for large $y$ and is not normalizable!

It may come as a surprise that both the
radion field (\reef{xxxx}) and the energy of the solution
(\reef{mass}) are linear in  the Schwarzschild mass parameter $\mu$.
Examining the effective lagrangian (3.71) of \cite{CGKP}, one sees
that it is quadratic in the radion, to leading order. Hence the
naive expectation would be that the Hamiltonian is also quadratic,
as opposed to linear, in $\mu$. Without going into any great detail,
the source of this `discrepancy' is the fact that the background
solution is time-dependent. That is, even though the bulk spacetime (\ref{minkowski})
has a Killing time\footnote{Of course, this is also the
asymptotic time conjugate to the energy which we measure with
$\mu$.}, the embedding of the standard SA brane does not
respect this symmetry. Given this time dependence, in
fact, we should in general expect that the Hamiltonian is linear in
perturbations about the background. This is most easily illustrated
with the simple example of a classical mechanics problem with ${\cal
H}=p\,\dot{q}-{\cal L}$. If we perturb about a specific solution
($p_0,q_0$), it is a straightforward calculation to show that in
general $\delta {\cal H}=\dot{q}_0\,\delta p - \dot{p}_0\,\delta q$.
Hence unless the background solution is static, we should expect the
shift in the energy to be linear in the perturbations. Of course,
the same analysis should apply directly to the present problem with
perturbations around the standard cosmology of SA brane (with $\mu=0$).

In summary we have confirmed our naive expectations: namely that a
Schwarzschild bulk corresponds to a normalisable radion on the SA
branch, and a {\it non}-normalizable radion on the N branch.

\end{document}